\documentclass[aps,prd,preprint,groupedaddress,showpacs]{revtex4}

\usepackage{epsfig}
\usepackage{graphics}
\usepackage{slashed}

\begin{document}

\preprint{DESY~10--025\hspace{12.4cm} ISSN 0418-9833}
\preprint{March 2010\hspace{15.55cm}}

\title{Inclusive $b$ and $b\bar b$ production with quasi-multi-Regge
kinematics at the Tevatron}

\author{\firstname{B.A.} \surname{Kniehl}}
\email{kniehl@desy.de}
\affiliation{{II.} Institut f\"ur Theoretische Physik, Universit\" at Hamburg,
Luruper Chaussee 149, 22761 Hamburg, Germany}

\author{\firstname{V.A.} \surname{Saleev}}
\email{saleev@ssu.samara.ru}
\author{\firstname{A.V.} \surname{Shipilova}}
\email{alexshipilova@ssu.samara.ru}
\affiliation{Samara State University, Academic Pavlov Street~1, 443011 Samara,
Russia}

\begin{abstract}
We consider $b$-jet hadroproduction in the
quasi-multi-Regge-kinematics approach based on the hypothesis of
gluon and quark Reggeization in $t$-channel exchanges at high
energies. The preliminary data on inclusive $b$-jet and $b\bar
b$-dijet production taken by the CDF Collaboration at the Fermilab
Tevatron are well described without adjusting parameters. We find
the main contribution to inclusive $b$-jet production to be the
scattering of a Reggeized gluon and a Reggeized $b$-quark to a $b$
quark, which is described by the effective Reggeon-Reggeon-quark
vertex. The main contribution to $b\bar b$-pair production arises
from the scattering of two Reggeized gluons to a $b\bar b$ pair,
which is described by the effective Reggeon-Reggeon-quark-quark
vertex. Our analysis is based on the Kimber-Martin-Ryskin
prescription for unintegrated gluon and quark distribution
functions using as input the Martin-Roberts-Stirling-Thorne
collinear parton distribution functions of the proton.
\end{abstract}


\maketitle \maketitle

\section{Introduction}
\label{sec:one}

The study of $b$-jet and $B$-meson production at high-energy colliders, such as
the Fermilab Tevatron and the CERN Large Hadron Collider, is of great interest
for the test of perturbative quantum chromodynamics (QCD).
The presence of a heavy $b$ quark, with mass $m_b\gg \Lambda_{\rm QCD}$, where
$\Lambda_{\rm QCD}$ is the asymptotic scale parameter of QCD, in such processes
guarantees a large momentum transfer even if the transverse momentum of the
produced $b$ quark is small.
Thus, the strong-coupling constant remains small in the processes
discussed here, $\alpha_s(m_b)\alt0.1$.
The study of $b$-jet production should provide a more direct way to investigate
gluon and quark interactions at small distances than that of $B$-meson
production because, in the former case, there is no need for additional
assumptions concerning the non-perturbative dynamics of the transition from a
$b$ quark to a $B$ meson \cite{Binnewies:1998vm}.

The total center-of-mass (CM) energy at the Tevatron, $\sqrt{S}=1.96$~TeV in
Run~II, sufficiently exceeds the scale $\mu$ of the relevant hard processes, so
that $\sqrt{S}\gg \mu \gg \Lambda_{\rm QCD}$.
In such a high-energy regime, the contributions to the production cross section
from subprocesses involving $t$-channel exchanges of partons (gluons and
quarks) may become dominant.
Then, the transverse momenta of the incoming partons and their off-shell
properties can no longer be neglected, and we deal with {\it Reggeized}
$t$-channel partons.
In this so-called quasi-multi-Regge kinematics (QMRK), the particles
(multi-Regge) or groups of particles (quasi-multi-Regge) produced in the
collision are strongly separated in rapidity.
In the case of inclusive $b$-jet production, this implies the following:
a single $b$ quark is produced in the central region of rapidity, while other
particles, including a $\bar b$ quark, are produced at large rapidities.
In the case of associated $b\bar b$-pair production in the central rapidity
region, we also assume that there are no other particles in this region, so
that the $b\bar b$ pair is considered as a quasi-multi-Regge pair of particles.
The QMRK approach \cite{QMRK} is particularly appropriate for this kind of
high-energy phenomenology.
It is based on an effective quantum field theory implemented with the
non-Abelian gauge-invariant action including fields of Reggeized gluons
(Reggeons) \cite{Lipatov95} and Reggeized quarks \cite{LipatoVyazovsky}.

In this paper, we apply the QMRK approach to various cross section
distributions of $b$-jet hadroproduction.
Specifically, we study the transverse-momentum distribution of single $b$-jet
production and, for $b\bar b$-dijet production, the distributions in the
leading-jet transverse energy, the dijet invariant mass, and the azimuthal
angle between the $b$ and $\bar b$ jets.
We compare our results with preliminary experimental data obtained by the
CDF Collaboration \cite{CDF1,CDF2}.

\section{Amplitudes}
\label{sec:two}

We first study inclusive single $b$-jet production in $p\bar p$ collisions,
working in the fixed-flavor-number scheme with $n_f=5$ active quark flavors.
To leading order (LO) in the QMRK approach, there is only one partonic
subprocess, namely
\begin{equation}
Q_b(q_1)+R(q_2)\to b(k), \label{equ:RQ}
\end{equation}
where $Q_b$ and $R$ are the Reggeized $b$ quark and gluon, respectively, and
the four-momenta are labeled as indicated in the parentheses.
As the modulus of the transverse momentum ${\vec k}_T$ of the $b$ quark,
$k_T\geq 32$~GeV \cite{CDF1,CDF2}, sufficiently exceeds its mass $m_b$, it is
justified to assume beauty to be an active flavor in the proton.
The effective vertex mediating subprocess~(\ref{equ:RQ}) is given by
\cite{LipatoVyazovsky}
\begin{equation}
C_{Q_bR}^b(q_1,q_2)=i\sqrt{4\pi\alpha_s}\,T^a\overline{u}(k)
\gamma^{(-)\mu}(q_1,q_2)\Pi_\mu^{(+)}(q_2)\label{eq:QRb},
\end{equation}
where $T^a$ are the generators of the color gauge group SU($N_c$) with $N_c=3$
for QCD, $a=1,\ldots,N_c^2-1$ is the color index of the Reggeized gluon,
$k=q_1+q_2$,
\begin{eqnarray}
\gamma^{(\pm)}_\mu(q,p)&=&\gamma_\mu+\slashed{q}\frac{n^{\pm}_\mu}{p^{\pm}},
\nonumber\\
\Pi^{(\pm)}_\mu(q)&=&-\frac{q^\mp n^\pm_\mu}{2\sqrt{-q^2}},
\end{eqnarray}
with $n_\mu^\pm=(1,0,0,\mp 1)$ in the CM frame and $q^{\pm}=q\cdot n^\pm$.
In the following, we put $q_{1,2}^\mu=x_{1,2}^\mu+(0,{\vec q}_{1,2T},0)$,
where $P_1$ and $P_2$ denote the four-momenta of the incoming proton and
antiproton, and ${\vec q}_{1T}$ and ${\vec q}_{2T}$ the transverse momenta of
the Reggeized $b$ quark and gluon, respectively.
We then have
${\vec k}_T^2={\vec q}_{1T}^2+{\vec q}_{2T}^2
+2|{\vec q}_{1T}||{\vec q}_{2T}|\cos \phi_{12}$,
where $\phi_{12}$ is the azimuthal angle enclosed between ${\vec q}_{1T}$ and
${\vec q}_{2T}$.
The squared amplitude of subprocess~(\ref{equ:RQ}) reads \cite{Saleev2008}:
\begin{equation}
\overline{|{\cal M}(Q_bR\to b)|^2}=\frac{2}{3}\pi \alpha_s {\vec k}_T^2.
\label{equ:ampRG}
\end{equation}

At next-to-leading order (NLO) in the QMRK approach, the main contribution to
inclusive $b$-quark production arises from the partonic subprocess
\begin{equation}
R(q_1)+R(q_2)\to b(k_1)+\bar b(k_2),
\label{equ:RRqq}
\end{equation}
where the $b$ and $\bar b$ quarks are produced close in rapidity.
The contributions due to the other NLO processes, $R+Q_b\to g+b$,
$Q_q+\bar Q_q\to b+\bar b$, and $Q_q(\bar Q_q)+Q_b\to q(\bar q)+b$
are suppressed because, in the small-$x$ region, the parton
distribution function (PDF) of the gluon greatly exceeds the
relevant quark PDFs.
Using the effective Feynman rules of the QMRK approach, the effective vertex
mediating subprocess~(\ref{equ:RRqq}) may be written in the following form:
\begin{eqnarray}
C_{RR}^{b\bar b}(q_1,q_2)&=&4\pi\alpha_s
\left[-\frac{1}{\hat s}f^{abc}T^c\overline{u}(k_1)\gamma_\mu v(k_2)
C_{RR}^{g,\mu}(q_1,q_2)\right.
\nonumber\\
&&{}+\frac{i}{\hat t}T^a T^b
\overline{u}(k_1)\gamma^\mu(\slashed k_1-\slashed q_1)\gamma^\nu v(k_2)
\Pi^{(-)}_\mu(q_1)\Pi^{(+)}_\nu(q_2)
\nonumber\\
&&{}+\left.\frac{i}{\hat u}T^b T^a
\overline{u}(k_1)\gamma^\nu(\slashed k_1-\slashed q_2)\gamma^\mu v(k_2)
\Pi^{(-)}_\mu(q_1)\Pi^{(+)}_\nu(q_2)\right],
\end{eqnarray}
where $\hat s = (q_1 + q_2)^2$, $\hat t = (q_1 - k_1)^2$, and
$\hat u = (q_2 - k_1)^2$ are the Mandelstam variables, $a$ and $b$ are the
color indices of the Reggeized gluons carrying the four-momenta $q_1$ and
$q_2$, respectively, and \cite{QMRK}
\begin{equation}
C_{RR}^{g,\mu}(q_1,q_2)=\frac{q_1^+q_2^-}{2\sqrt{q_1^2q_2^2}}
\left[(q_1-q_2)^\mu+\frac{(n^+)^\mu}{q_1^+}\left(q_2^2+q_1^+q_2^-\right)
-\frac{(n^-)^\mu}{q_2^-}\left(q_1^2+q_1^+q_2^-\right)\right]
\end{equation}
is the effective Reggeon-Reggeon-gluon vertex with the color structure stripped
off.
The squared amplitude of subprocess~(\ref{equ:RRqq}) was obtained
in Ref.~\cite{SaleevVasinBc}.
It may be presented as the linear combination of an Abelian and a non-Abelian
term, as
\begin{equation}
\overline{|{\cal M}(R + R \to b + \bar b)|^2} = 256 \pi^2
\alpha_s^2 \left[ \frac{1}{2 N_c} {\cal M}_{\mathrm{A}} +
\frac{N_c}{2 (N_c^2 - 1)} {\cal M}_{\mathrm{NA}}
\right],
\label{amp:RRbb}
\end{equation}
where
\begin{eqnarray}
{\cal M}_{\mathrm{A}}&=&\frac{t_1 t_2}{{\tilde t} {\tilde u}} -
\left( 1 + \frac{\alpha_1\beta_2 S}{\tilde
u}+\frac{\alpha_2\beta_1S}{\tilde t} \right)^2,
\nonumber\\
{\cal M}_{\mathrm{NA}} &=&
\frac{2}{S^2}\left(\frac{\alpha_1\beta_2S^2}{{\tilde
u}}+\frac{S}{2}+\frac{\Delta}{\hat
s}\right)\left(\frac{\alpha_2\beta_1S^2}{{\tilde
t}}+\frac{S}{2}-\frac{\Delta}{\hat s}\right)\nonumber\\
&&{}-\frac{t_1 t_2}{x_1x_2{\hat s}}\left[\left(\frac{1}{{\tilde
t}}-\frac{1}{{\tilde
u}}\right)(\alpha_1\beta_2-\alpha_2\beta_1)+\frac{x_1x_2 {\hat
s}}{{\tilde t} {\tilde u}}-\frac{2}{S}\right],
\nonumber\\
\Delta&=&\frac{S}{2}\left[{\tilde u} - {\tilde t}+2
S(\alpha_1\beta_2-\alpha_2\beta_1) +t_1
\frac{\beta_1-\beta_2}{\beta_1+\beta_2} -t_2
\frac{\alpha_1-\alpha_2}{\alpha_1+\alpha_2}\right],
\label{amp:RRbbDelta}
\end{eqnarray}
$\tilde t = \hat t - m^2$, $\tilde u = \hat u - m^2 $, $t_1 = -
q_1^2$, $t_2 = - q_2^2$, $\alpha_1=2(k_1\cdot P_2)/S$,
$\alpha_2=2(k_2\cdot P_2)/S$, $\beta_1=2(k_1\cdot P_1)/S$, and
$\beta_2=2(k_2\cdot P_1)/S$, with $S=(P_1+P_2)^2$.
To obtain the inclusive single $b$-jet production cross section, one needs to
integrate the cross section of subprocess~(\ref{equ:RRqq}) over the
$\bar b$-quark momentum.

At LO, $b\bar b$-dijet production receives contributions from both
subprocess~(\ref{equ:RRqq}) and the annihilation of a Reggeized quark-antiquark
pair,
\begin{equation}
Q_q(q_1)+\bar Q_q(q_2)\to b(k_1)+\bar b(k_2),
\label{eq:QQbb}
\end{equation}
where $q=u,d,s,c,b$.
Let us first consider the case $q\ne b$.
Neglecting Reggeized-quark masses, the effective vertex mediating 
subprocess~(\ref{eq:QQbb}) is given by \cite{LipatoVyazovsky}:
\begin{equation}
C_{Q_q\bar Q_q}^{b\bar b}(q_1,q_2,k_1,k_2)= \frac{4\pi\alpha_s}{\hat
s} T^a \bar u(k_1)\gamma^\mu v(k_2)\otimes T^a
\gamma_\mu^{(+-)}(q_1,q_2),
\end{equation}
where
\begin{equation}
\gamma_\mu^{(+-)}(q_1,q_2)=\gamma_\mu
-\frac{\slashed{q}_1n^{-}_{\mu}}{q_2^-}
-\frac{\slashed{q}_2n^{+}_{\mu}}{q_1^+}.
\end{equation}
The squared amplitude of subprocess~(\ref{eq:QQbb}) is found to be
\begin{equation}
\overline{|{\cal M}(Q_q\bar Q_q\to b\bar b)|^2}=
\frac{64\pi^2\alpha_s^2}{9x_1x_2\hat s^2}
\left(w_0+w_1S+w_2S^2\right) \label{eq:qqbb},
\end{equation}
where
\begin{eqnarray}
w_0&=&x_1x_2\hat s\left(\tilde t+\tilde u\right),\nonumber\\
w_1&=&-2x_2^2\alpha_1\alpha_2t_2-2x_1^2\beta_1\beta_2t_1+x_1x_2\{
(\alpha_2\beta_1+\alpha_1\beta_2)(\hat s+t_1+t_2)
+x_1x_2(\hat s-2m^2)
\nonumber\\
&&{}+x_1[\beta_1(t_1+\tilde u)+\beta_2(t_1+\tilde t)]
+x_2[\alpha_1(t_2+\tilde t)+\alpha_2(t_2+\tilde u)]\},
\nonumber \\
w_2&=&-2x_1x_2(\alpha_1\beta_2-\alpha_2\beta_1)^2.
\end{eqnarray}
In the case of $q=b$, we have
\begin{eqnarray}
C_{Q_b\bar Q_b}^{b\bar b}(q_1,q_2,k_1,k_2)&=&
4\pi\alpha_s\left[
\frac{1}{\hat s} T^a \bar u(k_1)\gamma^\mu v(k_2)\otimes
T^a \gamma_\mu^{(+-)}(q_1,q_2)\right.
\nonumber\\
&&{}+\left.\frac{1}{\hat t} T^a u(k_1)\gamma^{(-)\mu}(q_1,k_1-q_1)\otimes T^a
\gamma^{(+)}_\mu(-q_2,q_2-k_2)v(k_2)\right].
\end{eqnarray}
The analytic expression for $\overline{|{\cal M}(Q_b\bar Q_b\to
b\bar b)|^2}$ is too lengthy to be presented here.

\section{Cross sections}
\label{sec:three}

Exploiting the hypothesis of high-energy factorization, we may write the
hadronic cross sections $d\sigma$ as convolutions of partonic cross sections
$d\hat\sigma$ with unintegrated PDFs $\Phi_a^h$ of Reggeized partons $a$ in the
hadrons $h$.
For the processes under consideration here, we have
\begin{eqnarray}
d\sigma(p\bar p\to b X)&=&\int\frac{dx_1}{x_1}\int
\frac{d^2q_{1T}}{\pi}\int\frac{dx_2}{x_2}\int
\frac{d^2q_{2T}}{\pi} \left[\Phi^p_{b}(x_1,t_1,\mu^2)\Phi^{\bar
p}_{g}(x_2,t_2,\mu^2)\right.
\nonumber\\
&&{}+\left.\Phi^p_{g}(x_1,t_1,\mu^2)\Phi^{\bar p}_{b}
(x_2,t_2,\mu^2)\right]d\hat\sigma(Q_bR\to b),
\nonumber\\
d\sigma(p\bar p\to b\bar b X)&=&\int\frac{dx_1}{x_1}\int
\frac{d^2q_{1T}}{\pi}\int\frac{dx_2}{x_2}\int
\frac{d^2q_{2T}}{\pi}\left\{
\Phi^p_{g}(x_1,t_1,\mu^2)\Phi^{\bar p}_{g}(x_2,t_2,\mu^2)\right.
\nonumber\\
&&{}\times d\hat\sigma(RR\to b\bar b)
+\sum_q\left[\Phi^p_{q}(x_1,t_1,\mu^2)\Phi^{\bar p}_{\bar q}(x_2,t_2,\mu^2)
\right.
\nonumber\\
&&{}+\left.\left.
\Phi^p_{\bar q}(x_1,t_1,\mu^2)\Phi^{\bar p}_{q}(x_2,t_2,\mu^2)\right]
d\hat\sigma(Q_q\bar Q_q\to b\bar b)
\right\}.
\end{eqnarray}
The unintegrated PDFs $\Phi_a^h(x,t,\mu^2)$ are related to their collinear
counterparts $F_a^h(x,\mu^2)$ by the normalization condition
\begin{equation}
xF_a^h(x,\mu^2)=\int^{\mu^2}dt\,\Phi_a^h(x,t,\mu^2),
\end{equation}
which yields the correct transition from formulas in the QMRK approach to those
in the collinear parton model, where the transverse momenta of the partons are
neglected.
In our numerical analysis, we adopt the Kimber-Martin-Ryskin prescription
\cite{KMR} for unintegrated gluon and quark PDFs, using as input the
Martin-Roberts-Stirling-Thorne collinear PDFs of the proton \cite{MRST}.

For the reader's convenience, we collect here compact formulas for the
differential cross sections.
In the case of inclusive single $b$-jet production, we have \cite{Saleev2009}
\begin{eqnarray}
\frac{d\sigma}{dk_T\,dy}(p\bar p\to bX)&=&\frac{1}{k_T^3}\int
d\phi_1\int dt_1\left[\Phi^p_b(x_1,t_1,\mu^2)\Phi^{\bar
p}_g(x_2,t_2,\mu^2)\right.
\nonumber\\
&&{}+\left.\Phi^p_g(x_1,t_1,\mu^2)\Phi^{\bar
p}_b(x_2,t_2,\mu^2)\right] \overline{|{\cal M}(Q_bR\to b)|^2},
\label{eq:single}
\end{eqnarray}
where $y$ is the (pseudo)rapidity, $\phi_1$ is the azimuthal angle enclosed
between the vectors ${\vec q}_{1T}$ and ${\vec k}_T$,
\begin{equation}
x_{1,2}=\frac{k_T\exp(\pm y)}{\sqrt{S}},\qquad
t_2=t_1+k_T^2-2k_T\sqrt{t_1}\cos\phi_1.
\end{equation}
In the case of $b\bar b$-dijet production, we have
\begin{eqnarray}
\frac{d\sigma(p\bar p\to b\bar b X)}{dk_{1T}\,dy_1\,dk_{2T}\,dy_2\,d\Delta\phi}
&=&\frac{k_{1T}k_{2T}}{16\pi^3S^2}\int dt_1\int d\phi_1\frac{1}{(x_1x_2)^2}
\left\{\Phi^p_{g}(x_1,t_1,\mu^2)\Phi^{\bar p}_{g}(x_2,t_2,\mu^2)
\right.
\nonumber\\
&&{}\times\overline{|{\cal M}(RR\to b\bar b)|^2}
+\sum_q\left[\Phi^p_{q}(x_1,t_1,\mu^2)\Phi^{\bar p}_{\bar q}(x_2,t_2,\mu^2)
\right.
\nonumber\\
&&{}+\left.\left.
\Phi^p_{\bar q}(x_1,t_1,\mu^2)\Phi^{\bar p}_{q}(x_2,t_2,\mu^2)\right]
\overline{|{\cal M}(Q_q\bar Q_q\to b\bar b)|^2}\right\},
\label{eq:diff}
\end{eqnarray}
where $\phi_1$ is the azimuthal angle enclosed between the vectors
${\vec k}_{1T}$ and ${\vec q}_{1T}$, $\Delta\phi$ the one between
${\vec k}_{1T}$ and ${\vec k}_{2T}$,
\begin{eqnarray}
x_{1,2}&=&\frac{m_{1T}\exp(\pm y_1)+m_{2T}\exp(\pm y_2)}{\sqrt{S}},
\nonumber\\
m_{1,2T}&=&\sqrt{m^2+k_{1,2T}^2},
\nonumber\\
t_2&=&t_1+k_{1T}^2+k_{2T}^2+2k_{1T}k_{2T}\cos\Delta\phi
-2\sqrt{t_1}\left[k_{1T}\cos\phi_1+k_{2T}\cos(\Delta\phi-\phi_1)\right].
\end{eqnarray}
 In the massless limit, the $b$-quark transverse energy $E_{1T}$ is
given by $E_{1T}=k_{1T}$. The distribution in the $b\bar
b$-invariant mass $M_{b\bar b}$ may be easily obtained from
Eq.~(\ref{eq:diff}) by changing variables.

\section{Results}
\label{sec:four}

\begin{figure}[ht]
\begin{center}
\includegraphics[width=.8\textwidth, clip=]{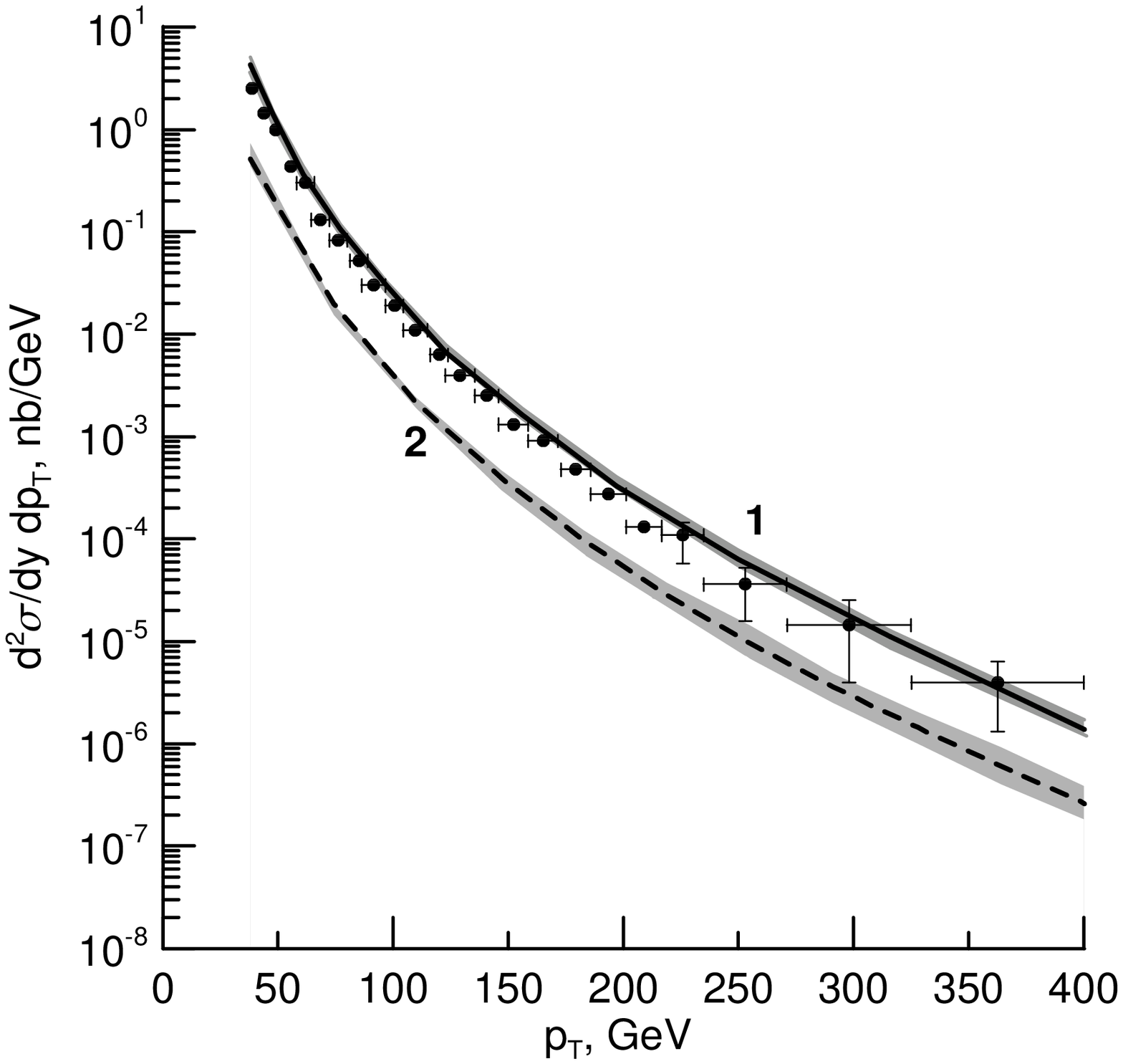}
\end{center}
\caption{\label{fig:1}%
The transverse-momentum distribution of inclusive single $b$-jet
hadroproduction measured by the CDF Collaboration at Tevatron Run~II
\cite{CDF1} is compared with the QMRK predictions due to
subprocesses~(\ref{equ:RQ}) 1 and (\ref{equ:RRqq}) 2.
The shaded bands indicate the theoretical uncertainties.}
\end{figure}

Recently, the CDF Collaboration presented preliminary data on
inclusive single $b$-jet production in $p\bar p$-collisions at
Tevatron Run~II \cite{CDF1}. The measurement was performed in the
kinematic range $38<k_T<400$~GeV and $|y|<0.7$. In
Fig.~\ref{fig:1}, these data are compared with our predictions
obtained in the QMRK approach as described in Secs.~\ref{sec:two}
and \ref{sec:three}. The contributions due to
subprocesses~(\ref{equ:RQ}) and (\ref{equ:RRqq}) are shown
separately. While the former may be evaluated from
Eq.~(\ref{eq:single}) as it stands, Eq.~(\ref{eq:diff}) must be
integrated over $k_{2T}$, $y_2$, and $\Delta\phi$ in the latter
case. Performing these integrations, care must be exercised to
avoid double counting, to ensure the separation of the $b$ jet
from the underlying event, and to guarantee infrared safety. In
the case of subprocess~(\ref{equ:RQ}), the $\bar b$ quark is
contained in the remnant of the hadron that emits the Reggeized
$b$ quark and is thus well separated from the final-state $b$
quark detected in the central region of the detector. In order to
avoid double counting, we therefore require for the $\bar b$ quark
of subprocess~(\ref{equ:RRqq}) to satisfy $|y_2|<4.5$. In fact,
the cross section due to subprocess~(\ref{equ:RRqq}) is negligibly
small for $|y_2|>4.5$, so that the precise value of this cut-off
is irrelevant. In order to implement the isolation of the $b$ jet,
we impose the acceptance cut $R_{\rm cone}>0.7$, where $R_{\rm
cone}=\sqrt{(y_1-y_2)^2+\Delta\phi^2}$, as in Ref.~\cite{CDF1}.
Since the lower bound of the $k_{2T}$ integration is zero, we
allow for the $b$-quark mass to be finite, $m_b=4.75$~GeV. The
renormalization and factorization scales are identified and chosen
to be $\mu=\xi k_T$, where $\xi$ is varied between 1/2 and 2 about
its default value 1 to estimate the theoretical uncertainty. The
resulting errors are indicated in Fig.~\ref{fig:1} as shaded
bands. We observe that the contribution due to
subprocess~(\ref{equ:RQ}) greatly exceeds the one due to
subprocess~(\ref{equ:RRqq}), by about one order of magnitude, and
practically exhausts the full result.
%The resulting error band is indicated in Fig.~\ref{fig:1} for the contribution
%due to subprocess~(\ref{equ:RQ}) only; it is very similar for the
%contribution due to subprocess~(\ref{equ:RRqq}).
%We observe that the former contribution greatly exceeds the latter, by about
%one order of magnitude, and practically exhausts the full result.
It nicely agrees with the CDF data throughout the entire $k_T$ range.
The QMRK results have to be taken with a grain of salt for $k_T\agt150$~GeV,
where the average values of the scaling variables $x_1$ and $x_2$ in the
unintegrated PDFs exceed 0.1, so that, strictly speaking, the QMRK approach
ceases to be valid.

\begin{figure}[ht]
\begin{center}
\includegraphics[width=.8\textwidth, clip=]{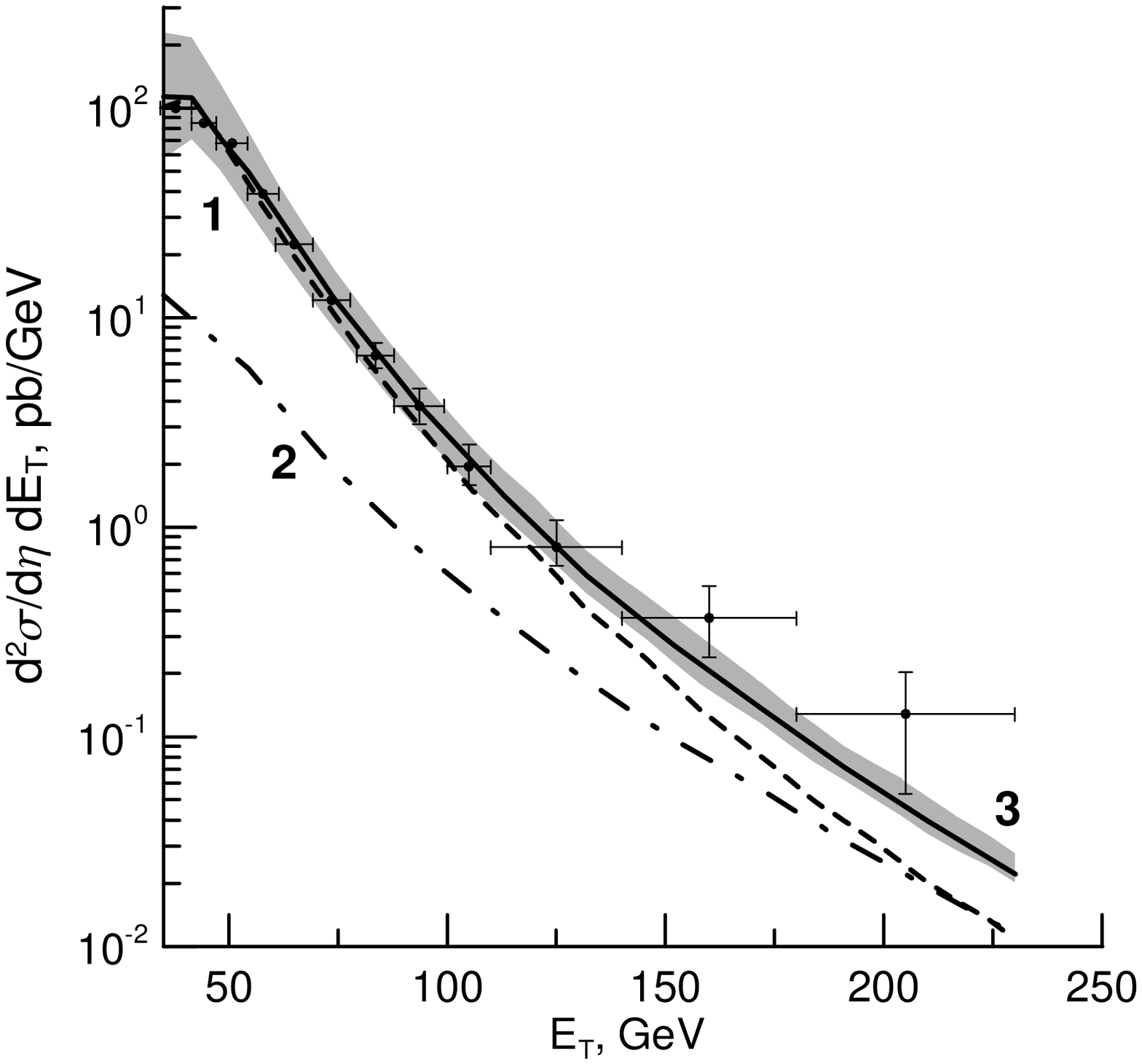}
\end{center}
\caption{\label{fig:2}%
The leading-jet transverse-energy distribution of inclusive $b\bar
b$-dijet hadroproduction measured by the CDF Collaboration at
Tevatron Run~II \cite{CDF2} is compared with the QMRK predictions
due to subprocesses~(\ref{equ:RRqq}) 1, (\ref{eq:QQbb}) 2, and their
sum 3.
The shaded band indicates the theoretical uncertainty on the latter.}
\end{figure}

\begin{figure}[ht]
\begin{center}
\includegraphics[width=.8\textwidth, clip=]{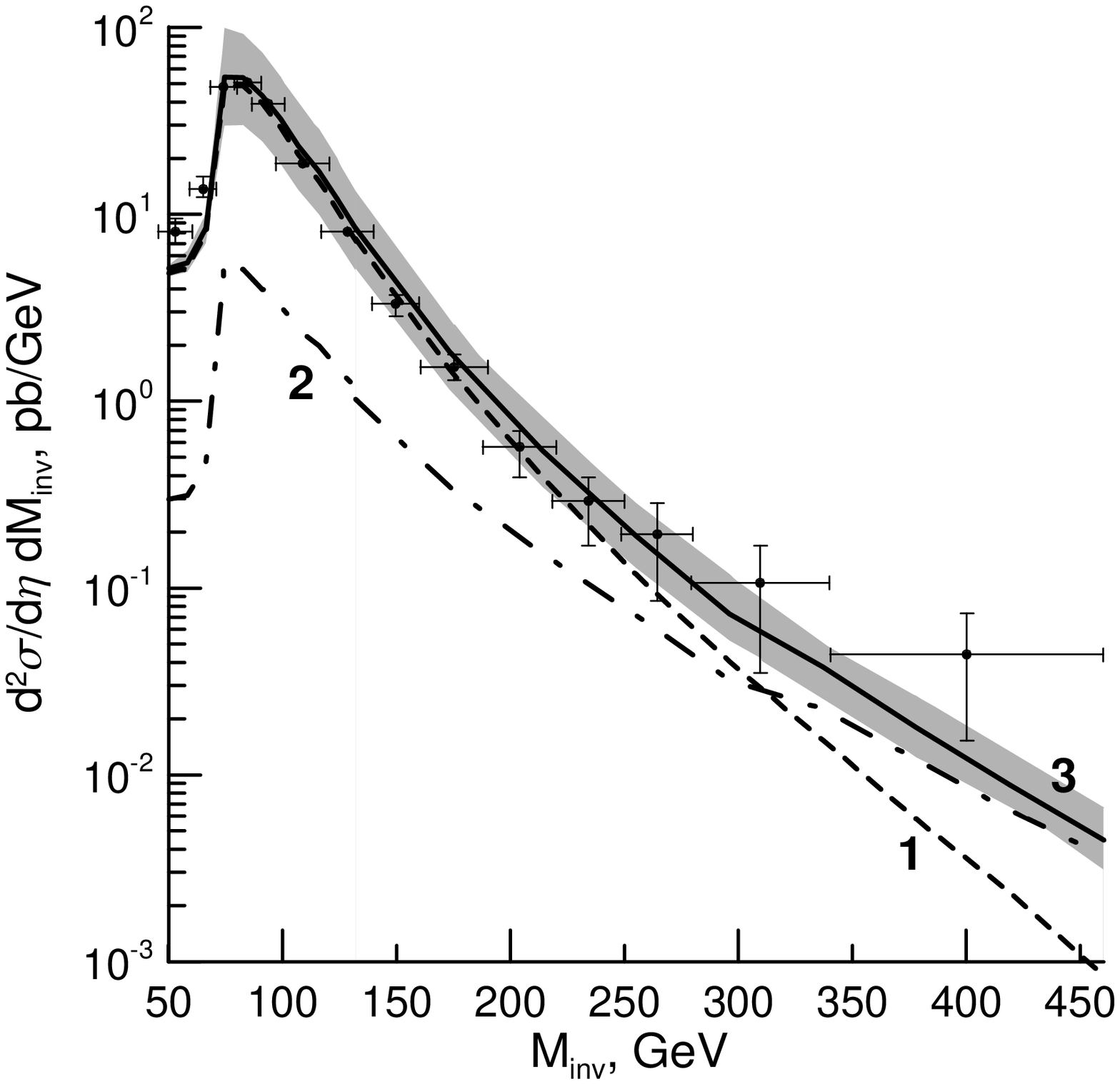}
\end{center}
\caption{\label{fig:3}%
The dijet-invariant-mass distribution of inclusive $b\bar b$-dijet
hadroproduction measured by the CDF Collaboration at Tevatron Run~II
\cite{CDF2} is compared with the QMRK predictions due to
subprocesses~(\ref{equ:RRqq}) 1, (\ref{eq:QQbb}) 2, and their sum 3.
The shaded band indicates the theoretical uncertainty on the latter.}
\end{figure}

\begin{figure}[ht]
\begin{center}
\includegraphics[width=.8\textwidth, clip=]{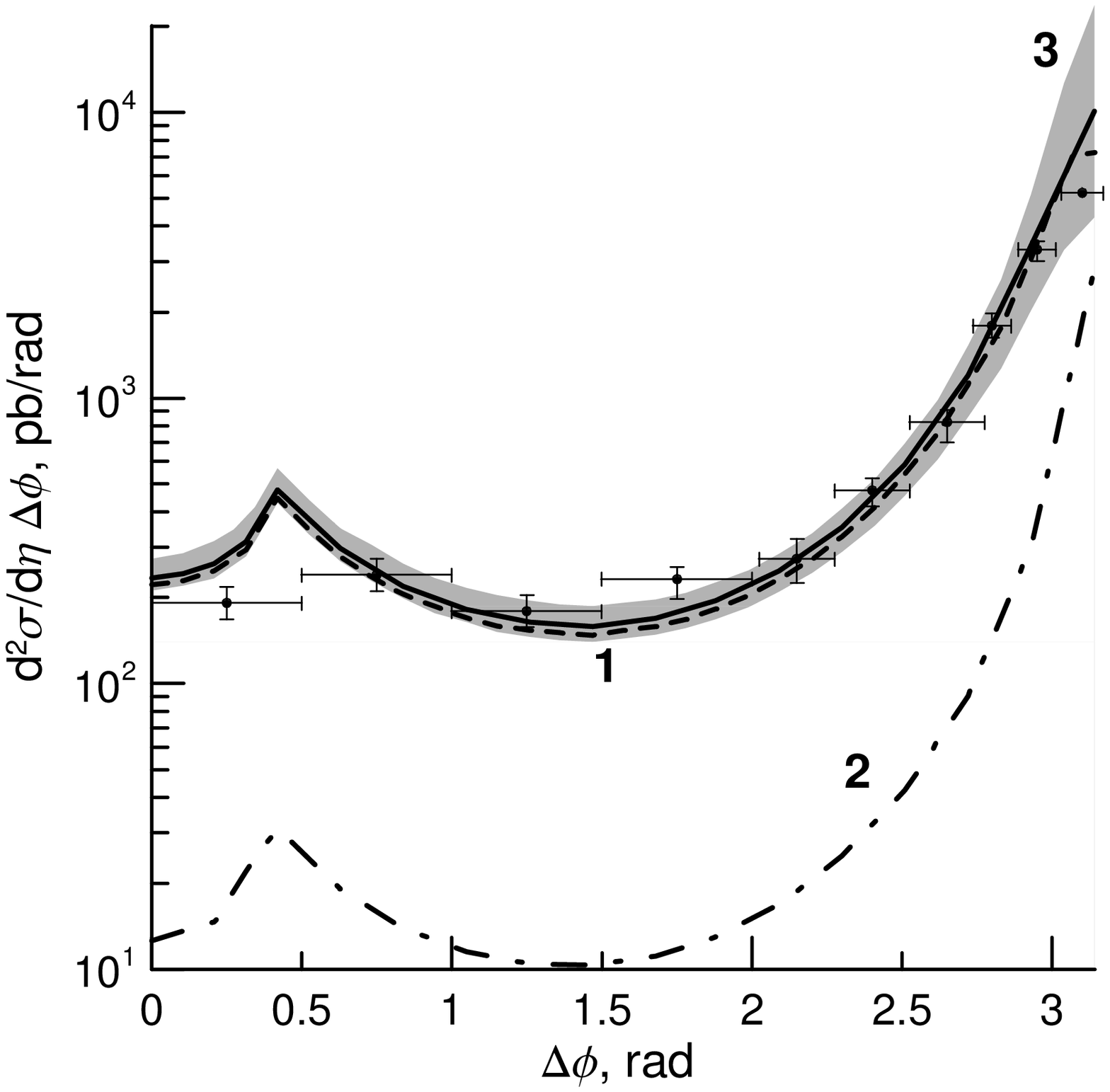}
\end{center}
\caption{\label{fig:4}%
The azimuthal-separation-angle distribution of inclusive $b\bar
b$-dijet hadroproduction measured by the CDF Collaboration at
Tevatron Run~II \cite{CDF2} is compared with the QMRK predictions
due to subprocesses~(\ref{equ:RRqq}) 1, (\ref{eq:QQbb}) 2, and their
sum 3.
The shaded band indicates the theoretical uncertainty on the latter.}
\end{figure}

The CDF Collaboration also measured the inclusive $b\bar b$-dijet
production cross section in Run~II at the Tevatron \cite{CDF2}. The
two jets were required to be in the central region of rapidity, with
$|y_1|,|y_2|<1.2$, to be separated by $R_{\rm cone}>0.4$, and to
have transverse energies satisfying the conditions $E_{1T}>35$~GeV
and $E_{2T}>32$~GeV, where the jet with the maximal transverse
energy is called the leading one. Given these acceptance cuts, the
massless approximation is clearly applicable, so that
$E_{iT}=k_{iT}$ and $y_i=\eta_i$, where $\eta_i$ denote the
pseudorapidities of the jets $i=1,2$. These data come as
distributions in the leading-jet transverse energy $E_{1T}$, the
dijet invariant mass $M_{b\bar b}$, and the azimuthal separation
angle $\Delta\phi$. They are compared with our QMRK predictions in
Figs.~\ref{fig:2}--\ref{fig:4}, respectively. The latter are
evaluated from Eq.~(\ref{eq:diff}) including the contributions from
subprocesses~(\ref{equ:RRqq}) and (\ref{eq:QQbb}). The common scale
is set to be $\mu=\xi k_{1T}$.
%{\bf or what else did you use?}.
In Figs.~\ref{fig:2}--\ref{fig:4}, these two contributions are shown separately
along with their superpositions.
The theoretical errors, estimated by varying $\xi$ between 1/2 and 2, are
indicated for the latter as shaded bands.
We observe that the total QMRK predictions nicely describe all the three
measured cross section distributions.
The contributions due to subprocess~(\ref{equ:RRqq}) dominate for
$E_{1T}\alt200$~GeV and $M_{b\bar b}\alt300$~GeV and over the whole
$\Delta\phi$ range considered.
The peak near $\Delta\phi=0.4$ in Fig.~\ref{fig:4} arises from the isolation
cone condition.

The shaded bands in Figs.~\ref{fig:1}--\ref{fig:4} only reflect the theoretical
errors due to the uncertainties in the choices of the renormalization and
factorization scales.
Additional and possibly larger errors arise from our lack of knowledge of the
unintegrated PDFs, which are, however, hard to quantify at this point.
%We thus leave a more detailed error analysis for future work. 

\section{Conclusions}
\label{sec:five}

We studied the inclusive hadroproduction of single $b$ jets and $b\bar b$
dijets at LO in the QMRK approach, including subprocesses~(\ref{equ:RQ}),
(\ref{equ:RRqq}), and (\ref{eq:QQbb}) with Reggeized partons in the initial
state.
Despite the great simplicity of our formulas, our theoretical predictions
turned out to describe recent measurements of various cross section
distributions by the CDF Colaboration in Run~II at the Tevatron surprisingly
well, without any ad-hoc adjustments of input parameters.
By contrast, in the collinear parton model of QCD, such a degree of agreement
can only be achieved by taking NLO corrections into account and performing
soft-gluon resummation.
In conclusion, the QMRK approach is once again \cite{KniehlSaleevShipilova}
proven to be a powerful tool for the theoretical description of QCD processes
in the high-energy limit.

\section*{Acknowledgements}

The work of B.A.K. was supported in part by the German Federal
Ministry for Education and Research BMBF through Grant No.\
05H09GUE, by the German Research Foundation DFG through Grant No.\
KN~365/7--1, and by the Helmholtz Association HGF through Grant No.\
HA~101. The work of V.A.S. and A.V.S. was supported in part by the
Federal Agency for Education of the Russian Federation under
Contract No.~P1338. The work of A.V.S. was also supported in part by
the International Center of Fundamental Physics in Moscow and the
Dynastiya Foundation.

\newpage

\end{document}